\documentclass[10pt,conference]{IEEEtran}

\usepackage[normalem]{ulem}
\usepackage{cprotect}
\usepackage{balance}
\usepackage{listings}
\usepackage{pdfpages}
\usepackage[normalem]{ulem}
\usepackage{listings}
\usepackage{enumitem}
\usepackage{graphicx}
\usepackage{amsmath}
\usepackage{xcolor}
\usepackage{colortbl}
\usepackage{makecell}
\usepackage{svg}

\usepackage{amssymb}
\usepackage{dsfont}
\usepackage{placeins}
\usepackage{algorithm}
\usepackage{algpseudocode}
\usepackage{tikz}
\usepackage{pifont}
\usepackage{microtype}
\usepackage{mdframed}
\usepackage{subcaption}
\usepackage{hyperref}
\usepackage{cleveref}
\usepackage{comment}
\usepackage{booktabs}

\usepackage{soul}
\soulregister\cite7
\soulregister\ref7
\soulregister\pageref7

\newtheorem{definition}{Definition}

\begin{document}
\title{Rewriting History: Repurposing Domain-Specific CGRAs}

\makeatletter
\newenvironment{btHighlight}[1][]
{\begingroup\tikzset{bt@Highlight@par/.style={#1}}\begin{lrbox}{\@tempboxa}}
	{\end{lrbox}\bt@HL@box[bt@Highlight@par]{\@tempboxa}\endgroup}

\newcommand\btHL[1][]{%
	\begin{btHighlight}[#1]\bgroup\aftergroup\bt@HL@endenv%
	}
	\def\bt@HL@endenv{%
	\end{btHighlight}%
	\egroup
}
\newcommand{\bt@HL@box}[2][]{%
	\tikz[#1]{%
		\pgfpathrectangle{\pgfpoint{1pt}{0pt}}{\pgfpoint{\wd #2}{\ht #2}}%
		\pgfusepath{use as bounding box}%
		\node[anchor=base west, fill=orange!30,outer sep=0pt,inner xsep=1pt, inner ysep=0pt, rounded corners=3pt, minimum height=\ht\strutbox+1pt,#1]{\raisebox{1pt}{\strut}\strut\usebox{#2}};
	}%
}
\newcommand{\name}[0]{FlexC}
\makeatother

\lstdefinestyle{C}{
	language={C},
	moredelim=**[is][{\btHL[fill=orange!30]}]{<ra>}{</ra>},
	moredelim=**[is][{\btHL[fill=gray!20]}]{<bi>}{</bi>},
	moredelim=**[is][{\btHL[fill=pink!20]}]{<pi>}{</pi>},
	moredelim=**[is][{\btHL[fill=green!20]}]{<gi>}{</gi>}
}

\author{
\IEEEauthorblockN{Jackson Woodruff}
\IEEEauthorblockA{University of Edinburgh \\ J.C.Woodruff@sms.ed.ac.uk} \\
\IEEEauthorblockN{Thomas K\oe hler}
\IEEEauthorblockA{INRIA \\ thomas.koehler@thok.eu}
\and
\IEEEauthorblockN{Alexander Brauckmann}
\IEEEauthorblockA{University of Edinburgh \\ alexander.brauckmann@ed.ac.uk} \\
\IEEEauthorblockN{Chris Cummins}
\IEEEauthorblockA{Meta AI Research \\ cummins@meta.com}
\and
\IEEEauthorblockN{Sam Ainsworth}
\IEEEauthorblockA{University of Edinburgh \\ sam.ainsworth@ed.ac.uk} \\
\IEEEauthorblockN{Michael F.P. O'Boyle}
\IEEEauthorblockA{University of Edinburgh \\ mob@inf.ed.ac.uk}
}
\date{}
%
%

%
%
%
\maketitle

\begin{abstract}


        Coarse-grained reconfigurable arrays (CGRAs) are domain-specific devices
	promising both the flexibility of FPGAs and the
	performance of ASICs.
	However, with  restricted domains comes a danger:
	designing chips that cannot accelerate enough current
	and future software to justify the hardware cost.

	We introduce \textit{\name{}}, the first flexible CGRA compiler, which allows
	CGRAs to be adapted to operations they do not natively support. 
	\name{} uses dataflow rewriting, replacing unsupported regions of
	code with equivalent operations that are supported by the CGRA\@.
	We use equality saturation, a technique enabling efficient exploration
	of a large space of rewrite rules, to effectively search through
	the program-space for supported programs.


        We applied  \name{} to over 2,000 loop kernels, compiling to four different research CGRAs and 300 generated CGRAs and demonstrate a 
        2.2$\times$ increase 
	in the number of loop kernels accelerated
	leading to 3$\times$ speedup compared to an Arm A5 CPU
 on kernels that would otherwise
	be unsupported by the accelerator.

\end{abstract}


\section{Introduction}

Specialized hardware has demonstrated truly significant performance gains
over general-purpose processors~\cite{Fuchs2019}, yet
 despite its potential~\cite{Taylor2012, IEEE2020},
 it faces  real  challenges to wider adoption~\cite{Dally2020}.
 The fundamental reason is that programming such accelerators is difficult~\cite{Corporaal2023}, often requiring modification
of the underlying algorithms~\cite{Dally2020}.
Users are often reluctant modify their algorithms~\cite{Silberstein2017} raising
frequency-of-use~\cite{Nowatzki2017,Brunvand2018,Gupta2022}
and cost~\cite{Khazraee2017} as concerns.

Heterogeneous Coarse-Grained Reconfigurable Architectures (CGRAs)~\cite{Liu2019a} are
a class of architectures that promise to solve this problem~\cite{Nowatzki2017}.
CGRAs can achieve near-ASIC level performance~\cite{Li2022a} and
provide enough flexibility to run a wider class of code~\cite{Nowatzki2017}.
Heterogeneous CGRAs use processing elements specialized
to various degrees~\cite{Aliagha2022}.
While specialization makes hardware more efficient~\cite{Bandara2022,VanEssen2011,Ebrahimi2021}, hardware specialization also introduces limitations on the software~\cite{Woodruff2022,Hooker2021,Woodruff2021}.

Despite aiming at flexibility, heterogeneous CGRAs are hard to use  
beyond  the scope they were designed for.
They age poorly as software evolves~\cite{Cong2013} and
falls out of the scope of the narrowly designed hardware: the
\emph{domain-restriction problem}.

This problem is highlighted by
existing state-of-the-art CGRA compilers 
such as OpenCGRA~\cite{tan2020} which
frequently fail
to generate code for the  
specialized hardware. If code contains even a single operation that
is unsupported by a particular hardware,
existing techniques simply cannot accelerate it,
restricting CGRAs to an overly narrow software domain.
This domain-restriction poses a significant challenge 
and is not
well understood~\cite{Wijvliet2016}.
What we need is a new approach that {\em automatically transforms} user programs to fit 
heterogeneous CGRAs, expanding the domain of supported software without user effort.


We introduce \textit{\name{}}, the first \textit{flexible} CGRA compiler that
addresses the domain-restriction problem.
\name{} uses a set of rewrite rules that translate unsupported operations into supported ones.
This compilation strategy requires a non-trivial application of
rewrites in an attempt to find a valid transformation 
to an expression the CGRA supports, leading to a 
large search space.  To explore this space
efficiently, \name{} uses a powerful technique called equality saturation~\cite{tate2009-equality-saturation,Willsey2021}.
CGRA compilation presents a number of unique
challenges to equality saturation including, crucially, transformation
encoding and cost modelling.
Overcoming these challenges
enables us to efficient explore large spaces.

In summary, we contribute:
\begin{itemize}[leftmargin=2.5em]
	\item \name{}, the first \textit{flexible} CGRA toolchain designed to support
		operationally-specialized CGRAs,
        increasing the number of loops that
        can run on a particular CGRA
        by a factor of 2.2$\times$;
	\item a compiler designed to integrate
            domain-specific rewrite rules, and four
            sets of rewrite rules demonstrating the effective
		translation of code to run on CGRAs designed for different domains;
	\item the first large-scale benchmark suite for CGRA compilers,
		with more than 2,000 loops from five different projects\footnote{To be released
		upon publication};
	\item an evaluation of these tools, demonstrating
		the importance of non-linear exploration
		techniques like equality saturation in finding
		working compilation sequences for real-world heterogeneous CGRAs.
\end{itemize}

\section{Motivation}

\begin{figure*}
	\includegraphics[width=\textwidth]{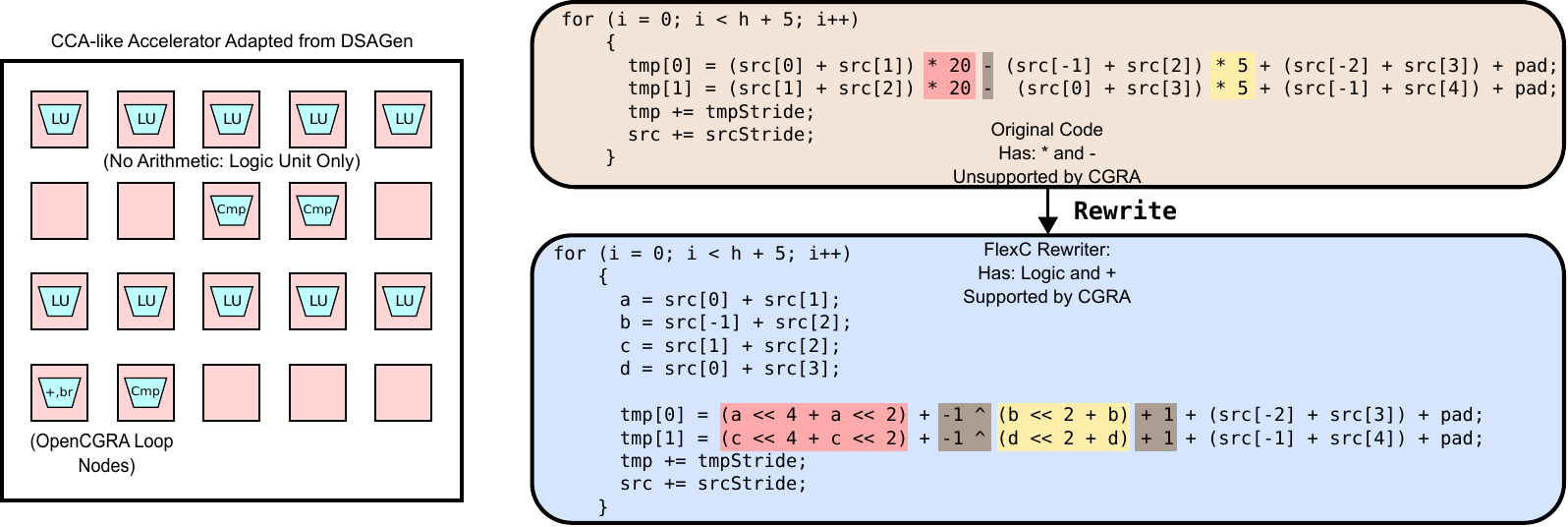}
		\vspace{-5pt}
	\cprotect\caption{An example from the FFMpeg~\cite{FFMpeg} library, which is part
		of our benchmark suite.
		\name{} rewrites the loop to run within the context of the
		CCA-like accelerator adapted from DSAGen~\cite{Weng2020}.
		Equality saturation is critical in this example to enable
		the conversion of \verb|a - b| into \verb|a +  -1 ^ b + 1|,
		as the rewriter must traverse the \verb|a + (-b)| state,
		which is no better than \verb|a - b|.  This is an example
		of the cost-trap problem (Figure~\ref{CostTrapProblem}).
	}
	\label{ExampleFFMPEG}
\end{figure*}
Heterogeneous CGRAs 
have the potential for 
better power efficiency and
lower area utilization than their 
homogeneous
counterparts~\cite{Bandara2022,Gobieski2021}. 
However, introducing this
heterogeneity introduces significant compilation challenges.

\subsection{The Software Domain-Restriction Problem}
 The cost of the specialized hardware has to be justified by 
enough use~\cite{Nowatzki2017} and
demand~\cite{Khazraee2017}.
To illustrate this problem, consider the loop  shown in figure~\ref{ExampleFFMPEG} from the FFMpeg library. We wish to compile this code  to the CCA-like accelerator adapted from DSAGen~\cite{Weng2020} shown on the left of the  figure. Unfortunately, the loop contains multiplication and subtraction operators which are not
supported by the CGRA. Currently, no compiler technique is able to generate code that is executable on this accelerator.
Our approach is able to rewrite the program into the form shown in the bottom of the figure. It no longer uses subtraction or multiplication, but instead 
used additions and shifts which are supported. The loop  can now  be executed on the CCA.

\subsection{Limits of existing compilers}
To overcome the domain-restriction problem, we need compilers to rewrite  
software that uses operations not natively supported by the hardware, but existing approaches fail.

\begin{figure}
	\begin{subfigure}{\columnwidth}
		\includegraphics[width=\columnwidth]{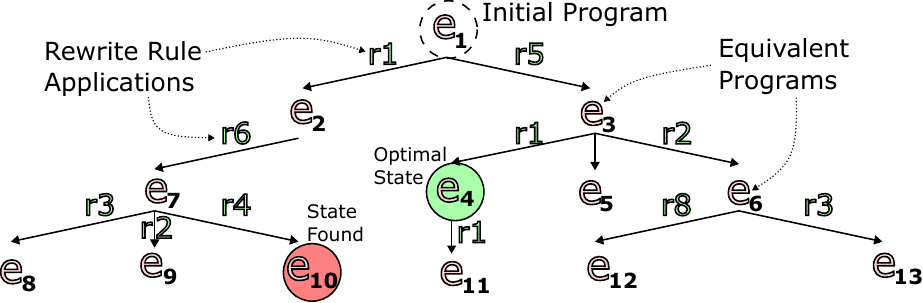}
		\caption{
			In the \textit{exploration problem},
			the expression in green
			is the optimal choice for this CGRA, but may never be reached
			in a greedy application of rewrite rules, which will reach the red
			state instead.
		}
		\label{ExplorationProblem}
	\end{subfigure}
	\begin{subfigure}{\columnwidth}
		\includegraphics[width=\columnwidth]{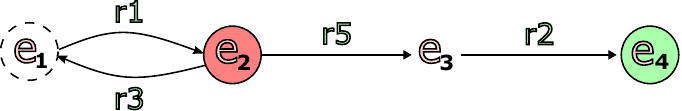}
		\caption{
			In the \textit{cycle problem},
			A greedy rewriter may get stuck in a cycle due to cyclical
			groups of rules, preventing it from finding
			the optimal state.
		}
		\label{RewriteRuleLoop}
	\end{subfigure}
	\begin{subfigure}{\columnwidth}
		\includegraphics[width=\columnwidth]{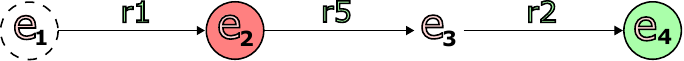}
		\caption{
			In the \textit{cost-trap problem},
			A greedy rewriter can get stuck in state e$_2$ as
			e$_{3}$ is a less valuable state.  
		}
		\label{CostTrapProblem}
	\end{subfigure}
	\caption{Applying rewrite rules with a greedy rewriter
		results in dead-ends that equality saturation
		avoids.}
	\label{RewritingProblems}
\end{figure}

\paragraph{Standard compiler flows}
Compiler frameworks, such as GCC, LLVM and MLIR, use canonicalization passes to transform IRs into a
predictable and efficient form.
Canonicalization is implemented with a set of simple fixed rewrite rules that are applied greedily.
In heterogeneous CGRAs, canonicalizing, however, does not solve the domain-restriction problem,
as the rewritten expressions may not be supported by the target hardware.
\paragraph{The Limits of Greedy Dataflow Rewriting}
For successful rewriting,
we need to add new rewrites
that allow translation to supported ones. 
However, we have to determine the order of, and whether to apply rewrites without searching a combinatorially large number of options. 
Greedy rewriting  is an efficient approach 
but Figure~\ref{RewritingProblems} highlights three problems that can cause greedy rewriters to get stuck.
In Figure~\ref{RewritingProblems}(a), greedily applying the first available rules, $r1,r6,r3$ to expression $e_{1}$ leads to the resulting expression $e_{10}$ which is less performant than the optimal expression $e_4$.
In Figure~\ref{RewritingProblems}(b), greedily applying the first available ruler  leads to a cycle  between  $e_1$ to $e_3$, never reaching the solution
$e_4$. Finally, in Figure~\ref{RewritingProblems}(c), the greedy rewriter gets stuck in a local minimum, $e_2$  due to the cost of applying further local rewrites.

Figure~\ref{ExampleFFMPEG} demonstrates these limitations in real code.  
To convert \verb|a - b| into \verb|a + -1 ^ b + 1|, the rewriter must traverse the \verb|a + (- b)| state, which is no better than \verb|a - b|.  As there is no immediate improvement in cost,
a greedy scheme would not apply such a rewrite. Equality saturation however,  applies this rewrite leading to  the transformed code executable on the accelerator.

\subsection{Our approach}
\name{} automatically adapts
the software, replacing unsupported operations via dataflow rewriting.
\name{} adaptively chooses between traditional greedy rewriting
techniques and equality saturation~\cite{tate2009-equality-saturation, Willsey2019}.
This overcomes challenges with traditional canonicalization and greedy techniques while retaining fast execution where possible. 
Equality saturation uses a graph data structure, called an e-graph, 
to record semantically equivalent programs while space-efficiently  representing their different syntactic program variations.
Rewrites are directly applied to this graph, rapidly growing the set of equivalent program variations.
Rewrites are either be applied until saturation is reached and no rewrites can be applied any more, or until a pre-defined rewrite goal is reached~\cite{Koehler2022}.

Our results confirm that equality saturation enables \name{} to compile more software to the CGRA.

\

\section{System Overview}
\name{} is implemented in
 OpenCGRA~\cite{tan2020}, a CGRA compiler intended to target
 heterogeneous CGRAs.
 Given an input DFG, \name{} explores sequences of rewrites to eliminate the operations that are not supported by a specialized architecture from the DFG.
 After rewriting the DFG, \name{} uses OpenCGRA to target the hardware.

\begin{figure}
	\includegraphics[width=0.5\textwidth]{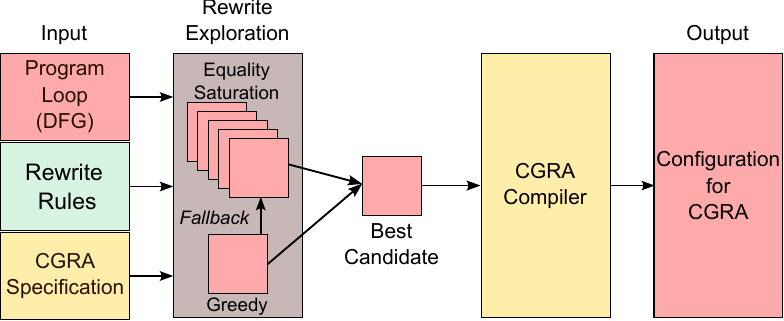}
	\vspace{2pt}
	\caption{
		\name{} system overview.  A Data-Flow Graph (DFG), set of Rewrite Rules and a CGRA
		specification are input.
		\name{} first applies a hybrid-rewrite strategy and selects the most
		suitable candidate to pass to
		the CGRA compiler, which generates the configuration.
	}
	\label{SystemOverviewFigure}
\end{figure}

Figure~\ref{SystemOverviewFigure} shows how \name{} compiles software
for a CGRA\@.  In a traditional CGRA compiler, a
Data-Flow Graph (DFG) is used to generate a CGRA configuration.
If the DFG does not match the target CGRA precisely, the code generation fails.

\name{} adds a rewrite system, using a set of rewrite
rules dictated by the context and a cost function based on the target CGRA\@.
After selecting the optimal graph according to the cost model
(the most likely DFG to be compilable to the underlying CGRA), \name{} uses
a traditional CGRA compiler to generate the final mapping.

\name{} can be applied in conjunction with any CGRA compiler --- provided
that appropriate rulesets using the right instructions can be supplied.
We provide \name{} under a liberal license to allow
this\footnote{Released upon publication}.

\newcommand{\rewritesTo}{\Rightarrow}

\subsection{Graph Rewriting}
\label{GraphRewritingSection}

\name{} translates programs to domain-specific CGRAs by generating a large set of equivalent code loops
and finding a suitable match if one exists.
This section formally defines our inputs: a dataflow graph representing a loop, a set of rewrite rules,
and a CGRA specification and our rewriting strategy.

\begin{definition}
A \emph{data flow graph} $D$ is a finite set of nodes $N$ corresponding to operations $op(n_1, ..., n_m)$, where $op$ is an operation symbol and $n_i \in N$ are children operands.

$D$ must be a \emph{directed acyclic graph}, meaning that a function $id: N \to |N|$ should exist such that:
$$\forall n = op(n_1, ..., n_m) \in N.~\forall i.~id(n_i) < id(n)$$
\end{definition}

While OpenCGRA uses Control Data Flow Graphs (CDFGs), and thus can handle branches and loops, we do not attempt to rewrite across control-flow boundaries. Instead, we break all control flow before rewriting, and restore control flow after rewriting.


\begin{definition}
A \emph{rewrite rule} $R$ is of the form $l \rewritesTo r$, where $l$ and $r$ are patterns.
A \emph{pattern} $P$ is a tuple $(N_P, O_P)$, where $N_P$ is a data flow graph that may contain variable nodes on top of operation nodes, and $O_P \subseteq N_P$ is a list of output nodes.

$R$ can be applied to $D$ when $l$ has a match $(M_l, \sigma)$ in $D$, where $M_l$ is a list of nodes from $D$ matching $O_l$, and $\sigma$ maps variables to matching nodes from $D$.
To produce the list of nodes $M_r$ that should replace the $M_l$ nodes in $D$, the variables are substituted in $r$, written as $r[\sigma] = (N_r', M_r)$.
\end{definition}

A rewrite rule must be \emph{semantics-preserving}.
This means that, $\forall (M_l, \sigma).~ M_l = M_r$ which
depends on the element-wise application of a given semantic equality.
The meaning of equality in this case depends on the rules provided.  We will
see in section~\ref{RewriteRules} that this may be true equality, fuzzy equality (e.g.\ with floating-point
manipulation rules) or even weaker definitions of equality
(e.g.\ with stochastic computing rules~\ref{stochasticcomputing}).


\begin{definition}
In a CGRA, we have an array of processing elements, PEs ($\mathit{PE}_i$), each
of which supports a particular set of operations ($op(n_1, ..., n_m)$), $\mathit{Supported}_i$. We generate this set from the CGRA's specification.

Given a particular DFG $D$, with nodes $N$, there may be some subset of
nodes $\mathit{Unsupported}(N)$ that have operations without
hardware support \emph{anywhere} on the CGRA\@.  We wish to find a sequences of rewrite
rules that we can apply to the DFG to produce $D'$ with nodes $N'$
such that $\mathit{Unsupported}(N') = \{\}$, as otherwise it will be impossible
to schedule that particular code onto the
CGRA\@.  We thus define the set of operations a particular full CGRA
can support as:
$$ops = \bigcup_{i} \mathit{Supported}_i$$
\end{definition}






\subsubsection{Rewriting Goal}
\label{rewriting-goal}
\label{CostModelSection}

The compiler takes a dataflow graph (DFG) $D$ as input.  Numerous existing
techniques attempt to find a valid mapping~\cite{Martin2022}, but in heterogeneous CGRAs,
the operations in the DFG may not be in the supported set for \textit{any} node.

The goal of a rewriting algorithm $A(D, Rs, ops)$ is to return $D'$, obtained by rewriting $D$ using the set of rules $Rs$, such that $D'$ only uses operation symbols from $ops$. 

We further define a cost function $C(D, ops)$ to minimize:
$$ \sum_{op(...) \in N} 1 \text{ if } op \in ops \text{ else } 10^6 $$
This incentivizes smaller programs by giving a cost of $1$ to available operations,
while giving a huge penalty to unavailable operations by giving them a cost of $10^6$.
Our CGRA specification and cost function aim to eliminate
unavailable operations to successfully map the program onto the CGRA, without
trying to precisely model the execution performance.

With the assumption that $|N| < 10^6$, rewriting successfully
eliminates all unavailable operations if $C(D', ops) < 10^6$, and
fails to do so if $C(D', ops) \geq 10^6$.

\subsubsection{Greedy Rewriting}
\Cref{alg:greedy-rewriting} shows our greedy rewriter.
Greedy rewriting is the most straightforward rewriting approach; it
runs quickly but often gets stuck in local minima.

On each greedy iteration, we iterate over every
rewrite rule to find matches (lines \ref{l:greedy-match-beg} to \ref{l:greedy-match-end}).
If applying a rewrite for a given match leads to a cost reduction, we proceed with
the rewritten program and forget about the previous program
(lines \ref{l:greedy-apply-beg} to \ref{l:greedy-apply-end}).
The \lstinline{local_minima} variable keeps track of whether a fixed point was
reached, which is the termination condition (line \ref{l:greedy-loop}).

\begin{lstlisting}[float, label={alg:greedy-rewriting}, caption={Greedy rewriting algorithm}, language=python, frame=tb, numbers=left, xleftmargin=2em, mathescape,basicstyle=\small]
def greedy(d, rs, ops):
  local_minima = false
  while not local_minima:$\label{l:greedy-loop}$
    local_minima = true

    for r in rs:$\label{l:greedy-match-beg}$
      matches = find_matches(d, r)
      for m in matches:$\label{l:greedy-match-end}$
          d2 = apply_match(d, m)$\label{l:greedy-apply-beg}$
          if C(d2, ops) < C(d, ops):
            d = d2$\label{l:greedy-apply-end}$
            local_minima = false
            break
      
  return d
\end{lstlisting}



\subsubsection{Equality Saturation}
\label{equality-saturation}

\Cref{alg:equality-saturation} shows our algorithm for rewriting via equality saturation.
Equality saturation~\cite{tate2009-equality-saturation} is a more sophisticated rewriting approach;
it avoids getting stuck in local minima but can be slow to execute.
We leverage both the state-of-the-art Rust \texttt{egg} library~\cite{Willsey2021} and existing
work extending equality saturation to graph rewriting~\cite{yang2021equality}.

First, we initialize an e-graph data structure that compactly represents a space of equivalent
programs by sharing equivalent sub-terms as much as possible (line \ref{l:eqsat-init}).
Then, we run the explorative phase of equality saturation using our set of rewrite rules,
iteratively exploring possible rewrites in a breadth-first manner and growing the
e-graph (line \ref{l:eqsat-explore}).

As visible in line \ref{l:eqsat-loop}, the explorative phase
terminates when all possible rewrites have been explored (a fixed point, called saturation, is reached),
or when another stopping criteria is reached (e.g.\ a timeout).
On each explorative iteration, all rewrite-rule matches are
collected (line \ref{l:eqsat-match}) and applied in a non-destructive way,
adding new equalities into the e-graph (line \ref{l:eqsat-apply}).

Finally, we extract the best program from the e-graph according to our cost
function using Linear Programming (line \ref{l:eqsat-extract}).

\begin{lstlisting}[float, label={alg:equality-saturation}, caption={Equality saturation algorithm}, language=python, frame=tb, numbers=left, xleftmargin=2em, mathescape,basicstyle=\small]
def eqsat(d, rs, ops):
  egraph = initialize_egraph(d)$\label{l:eqsat-init}$
  egg_exploration(egraph, rs)$\label{l:eqsat-explore}$
  return egg_lp_extraction($\label{l:eqsat-extract}$egraph,
                           cost_for_egg(ops))$\label{l:eqsat-cost}$

def egg_exploration(eg, rs):
    ...
    while not saturation_or_timeout:$\label{l:eqsat-loop}$
        matches = []
        for r in rs:
            matches += find_matches(eg, r)$\label{l:eqsat-match}$
        for m in matches:
            apply_match(eg, m)$\label{l:eqsat-apply}$
    ...
\end{lstlisting}


\subsubsection{Hybrid Rewriting}
\label{HybridRewritingSection}

\begin{lstlisting}[float, label={alg:hybrid-rewriting}, caption={Hybrid rewriting algorithm}, language=python, frame=tb, numbers=left, xleftmargin=2em, mathescape,basicstyle=\small]
def hybrid(d, rs, ops):
    d2 = greedy(d, rs, ops)
    if cost(d2, ops) < $10^6$: return d2
    else:                  eqsat(d, rs, ops)
\end{lstlisting}

\name{} uses hybrid rewriting (\cref{alg:hybrid-rewriting}), which takes the best
from both strategies.  In hybrid rewriting, we first
apply a fast greedy rewriter.  If the greedy rewriter
does not find a suitable candidate, \name{} falls back
to the more expensive, but more likely to succeed,
equality saturation.

\section{Rewrite Rules}
\label{RewriteRules}
\label{RewriteRulesSection}
\name{} is a platform that can target any domain-specific
CGRA, and can integrate domain-specific rules to work alongside
traditional rules.  Equality-saturation enables this flexibility
by using the same rule exploration algorithms regardless of
changes in the ruleset.
We explore several different rulesets: some rules
are always correct, while other rulesets may only be useful
in certain domains, such as the stochastic-computing rewrite rules (\cref{ssec:stochastic}).

In a traditional application of rewrite rules, compilers look
to perform strength reduction~\cite{Cooper2001}, by replacing
more complex rules with simpler rules --- this is typically
achieved by canonicalizing towards the simplest method of representing
an expression.
In a traditional compiler, a rule is typically formatted as:
\begin{align*}
	\text{Complex Operation} \rightarrow \text{Simpler Operation}
\end{align*}

A typical rewrite system produces a series of independent rewrites,\[e_1 \underset{\text{rule} 1}{\implies} \cdots \underset{\text{rule} N-1}{\implies} e_N\] to
produce the best suited expression.
The rules are written in such a way that
they chain together,
as they are in existing compilers.
We stop rewriting when no more rules are applicable.

However, when compiling for a CGRA, replacing simpler operations
with \textit{more complex}
ones can be beneficial if they enable an entire
region of code to be run on faster, fixed-function hardware.

As a result,
for some sequence of rules
\[e_1 \underset{\text{rule} 1}{\implies} \cdots \underset{\text{rule} i - 1}{\implies}  e_i \underset{\text{rule} i}{\implies}  \cdots \underset{\text{rule} N - 1}{\implies}  e_N\]
some intermediate $e_i$ may be the best choice of expression, and further, rule application can occur bidirectionally.
Rather than strength reduction, which implies a linear sequence of operations that become strictly simpler, the process for compiling for a CGRA is instead \textit{rewrite exploration}.


\subsection{Core Integer Rules}
\label{RulesetsSection}

We use a set of strength-reduction and canonicalization
rules representative of those in a typical compiler.
An example is:
\begin{lstlisting}[xleftmargin=.175\textwidth]
x * -1 => -x
\end{lstlisting}

On the left-hand side of this rule, we require a multiplication
operation, and on the right-hand side, we require
a negation operation.  For most compilers, the right-hand side
is (almost) always the better choice, so most
rewriters only apply these rules forwards

\name{} applies this rule in both directions, as some CGRAs may have
multiplication-supporting PEs and other CGRAs may have
negation-supporting PEs.  We refer to this universally applicable ruleset as the \textit{integer ruleset}. Some examples are shown in
Table~\ref{RewriteRulesTable}.

\subsection{Domain-Specific Rules}

The core integer ruleset represents a
baseline of rules widely applicable to all CGRAs.
However, the constrained nature of heterogeneous CGRAs, which may feature highly specialized
operations, means that domain-specific relaxations of correctness typically
result in better targetability.
\name{} can use custom rewrite rules as input, tailored for a given accelerator. 

\subsubsection{Floating-Point Rules}
Floating-point rewrite rules are rarely bit-for-bit correct.
Compilers typically use flags to allow for different levels
of correctness guarantees, enabling
floating-point transformations only when the programmer is willing
to forgo accuracy.

When compiling floating-point operations to CGRAs, 
\name{} uses these rules by default (they can be turned off).
This enables more rewrites at the cost of losing bit-correctness.
An example of rules enabled by this assumption are shown in
Table~\ref{FPRewriteRulesTable}.

\subsubsection{Boolean Logical Operations}
Logical operations such as \verb|AND| (\&) and \verb|OR| (|) take two different
meanings: they specify bitwise operations
on entire words at a time, and they specify boolean operators (where any non-zero result is \texttt{true}).
With a compiler flag provided by a programmer to indicate these are equivalent,
we can add more rewrite rules.

For example, as boolean operations, \verb|AND| can be rewritten using multiplication
nodes, increasing the space of programs that a CGRA without logical operator
support can be used for.  We supply a set of rewrite operations that assume
logical operations are equivalent to boolean operations.
Some examples of rewrite rules in this set are shown in
Table~\ref{BooleanRewriteRulesTable}.

\begin{table}
	\centering
	\begin{tabular}{rcl}
		\verb|x - y|&\verb|<=>|&\verb|x + (-y)| \\
		\verb|x >> y|&\verb|<=>|&\verb|x / (1 << y)| \\
		\verb|x and y|&\verb|<=>|&\verb|not ((not x) or (not y))|
	\end{tabular}
	\caption{Some example rewrite rules that can be used to change
	the operations an expression requires.  }
\label{RewriteRulesTable}

			\centering
	\begin{tabular}{rcl}
		\verb|x * y| & \verb|<=>| & \verb|x / (1.0 / y)| \\
		\verb|-1.0 * x| & \verb|<=>| & \verb|-x |
	\end{tabular}
\caption{Rewrite rules enabled by reducing requirements
	on floating-point equality. }
\label{FPRewriteRulesTable}

		\centering
	\begin{tabular}{rcl}
		\verb|x AND y| & \verb|=>| & \verb|x * y| \\
		\verb|x OR y| & \verb|=>| & \verb|(x + y) > 0| \\
		\verb|x XOR y| & \verb|=>| & \verb|x != y| \\
	\end{tabular}
	\caption{Rewrite rules under the assumption that
	binary logical operators are boolean operators.
}
\label{BooleanRewriteRulesTable}
		\centering
	\begin{tabular}{rcl}
		\verb|x * y| & \verb|=>| & \verb|x AND y| \\
	\end{tabular}
	\caption{Example rewrite rule for stochastic computing}
\label{StochasticComputingRewriteRules}
\label{StochasticComputingRulesTable}

\end{table}

\subsubsection{Stochastic Computing}
\label{ssec:stochastic}
\label{stochasticcomputing}
Stochastic computing is a computing paradigm aimed at achieving
better energy efficiency than traditional computing by trading off
accuracy~\cite{Alaghi2013}.  In particular, stochastic
computing allows multipliers to be replaced by logical
\texttt{and} operators, and \texttt{add} operations to be replaced
by muxes~\cite{Baker2020}, in contexts where the absolute result is not needed, and this allows the use of simpler accelerators.
Table~\ref{StochasticComputingRulesTable} shows an example
ruleset.

\section{Experimental Setup}
We implement \name{} above OpenCGRA, which is written in C++.
We use the \texttt{egg} Rust library~\cite{Willsey2021} to implement our rewriters.  
For equality saturation, we use an iteration limit of 10 with a node limit of 100,000
to prevent the e-graphs from growing too large.
\name{} is integrated into the LLVM framework and is invoked using
the \texttt{opt} tool using LLVM IR as input.

\name{} relies on OpenCGRA to find the loop to accelerate.
OpenCGRA looks for the first loop in each provided function.
We implement the architecture specification in JSON,
adding a mapping from each PE to the sets of operations
it will be able to support.

\subsection{Benchmarks}
\begin{table}
  \centering
  \begin{tabular}{llr}
    \toprule
    Domain & Project & Samples \\
    \midrule
	Compression & Bzip2~\cite{BZip2} & 13\\
	Multimedia & FFmpeg~\cite{FFMpeg} & 1852\\
			   & FreeImage~\cite{FreeImage} & 223\\
	Scientific Computing & DarkNet~\cite{DarkNet} & 77\\
						 & LivermoreC~\cite{LivermoreC} & 26\\
    \midrule
     Sum &  & 2188\\
    \bottomrule
  \end{tabular}
  \caption{Quantities of unique loops in benchmark suite.}
\label{BenchmarkSuiteTable}
\end{table}

We have collected a benchmark suite of 2188 real-world open-source code loops
composed of projects in the multimedia, compression and simulation domains
shown in table~\ref{BenchmarkSuiteTable}.
Typically, CGRA compilers are evaluated on benchmark suites of a few tens
of loops  which do not capture the wide
spectrum of loops that programmers write, and are easy
to hand optimize~\cite{Intel-a}.  Our benchmark suite
captures a wide range of loops, without the overheads of running whole
programs~\cite{Arm2021}.
These loops allow us to demonstrate \name{} works on a wide range of architectures
and programs.

We extract loops suitable for CGRA scheduling from the projects shown in table~\ref{BenchmarkSuiteTable}.
Each extracted is the innermost loop, has no internal branches or function calls, and contains at least
one array access.
These properties ensure our benchmarks
compile to many different CGRAs using
a variety of compilation techniques.

We build a custom Clang-based tool that identifies loop structures and detects
required type definitions.  Clang is run using the build-system rules
for each project.
Each loop is placed into a function skeleton so it can be compiled.

\subsection{Alternative approaches}
We compare FlexC against three alternative
approaches: OpenCGRA~\cite{tan2020}, the LLVM~\cite{llvm} Rewriter and our own Greedy Rewriter.  OpenCGRA is
the default scheme that simply maps operations to function
units without any rewriting, with LLVM's rewriter 
disabled to enable a comparison to it.
The LLVM rewriter employs the rewrite rules
within the LLVM compiler infrastructure,
which are intended to cannonicalize the program
for typical CGRA architecture.
The greedy rewriter is FlexC without equality saturation
fallback.

\subsection{Existing Domain-Specific Accelerators}
\label{ExistingDSASection}
We evaluate domain-specific architectures from
three prior works.  We consider one domain-specific CGRA work (REVAMP~\cite{Bandara2022}),
one more general domain-specific \textit{accelerator} work (DSAGen~\cite{Weng2020}), and one
stochastic computing CGRA (SC-CGRA~\cite{Wang2022}).

\subsubsection{DSAGen}
DSAGen~\cite{Weng2020} is a framework for generating domain-specific
architectures.  These architectures share many properties
with CGRAs in that they expose architectural details to the compiler
and present coarse-grained reconfigurable blocks.
We make minor modifications\footnote{
	OpenCGRA requires more routing to be present
	between compute elements, so the architectures
	we use are more flexible than 
	in DSAGen.  OpenCGRA also does not support
	architectural features like distribution trees,
	which we have omitted.
	We further add the nodes required by OpenCGRA
	to support loop pipelining (an \texttt{add} and an integer
	compare).
} to the architectures
shown in Figure 4(b) and 4(c) in~\cite{Weng2020} so
they can be represented within OpenCGRA\@. 

\subsubsection{REVAMP}
REVAMP~\cite{Bandara2022}, a framework for
generating domain-specific CGRAs provides an example of a CGRA
for heterogeneous compute optimization, with nodes for addition,
subtraction, multiplication and some logic operations implemented
within a 6x6 CGRA\@.

\subsubsection{SC-CGRA}
SC-CGRA~\cite{Wang2022} is a stochastic-computing-based CGRA\@.  Typical exact multipliers
are replaced with approximate multipliers, and
similarly for adders within a 4x4 CGRA\@.
We implement this in OpenCGRA, providing
approximate adders/multiplers instead of exact ones\footnote{
The authors discuss different accuracies of adder/multiplier,
but do not state the number of each used, so we
use a simple assignment of one multiplier and
one adder per node.  We also omit node-fusing,
as we use OpenCGRA to target this accelerator.
The operators other than the multipliers and adders
are not specified completely.  For this evaluation,
we assume each node has logical operations, and
arithmetic operations simpler than multiplication.
To enable OpenCGRA to compile some things on it's own,
we add one exact adder, which is required for induction
variables in almost all loops.
}.  

\section{Results}
\label{results}
We evaluate \name{} against traditional heterogeneous-CGRA compilers, improving the number of
benchmarks that can be compiled to heterogeneous CGRAs
by 2.2$\times$, and demonstrate that despite making the computation
more complex, the rewrite rules do not introduce slow-downs,
showing geomean speedups of 3$\times$.

\subsection{Existing Domain-Specific Accelerators: Compilation Rate}
\begin{figure}
	\includegraphics[width=\columnwidth]{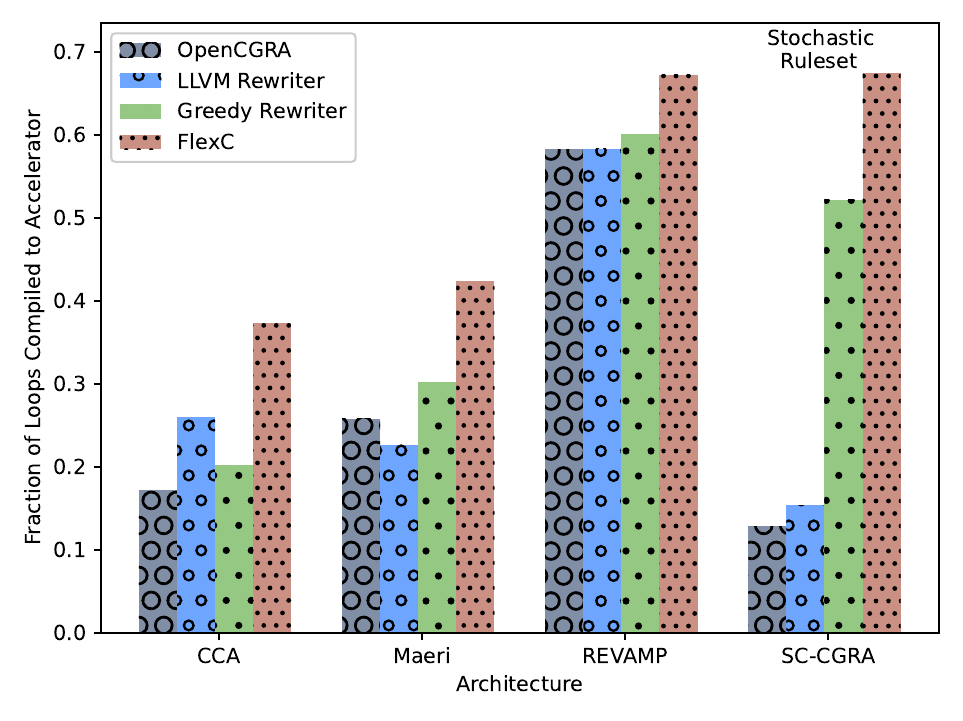}
	\caption{
		We consider four different architectures, adapted
		from DSAGen~\cite{Weng2020} (CCA and Maeri), REVAMP~\cite{Bandara2022}
		and SC-CGRA~\cite{Wang2022}.
		All architectures use the integer and floating-point rulesets,
		and SC-CGRA uses the stochastic ruleset.
		The architectures are specialized to different degrees:
		the more specialized architectures, CCA and Maeri,
		benefit from \name{} more than the more generic
		architecture from REVAMP\@.
	}
	\label{CaseStudiesCompilationRateGraph}
\end{figure}

\begin{figure*}
	\includegraphics[width=\textwidth]{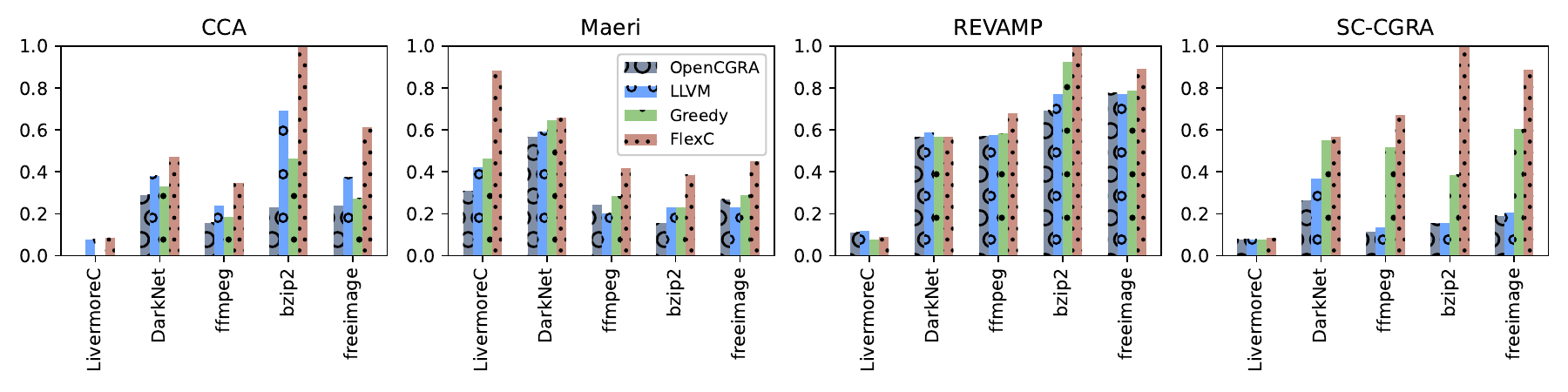}
	\caption{
		Results for each accelerator pairing by benchmark suite. Equality saturation often dramatically improves coverage for particular workload-accelerator combinations (e.g.~bzip2 on CCA and SC-CGRA, and LivermoreC on Maeri), where otherwise the accelerator would appear entirely unsuitable.
		In these cases, the accelerator has the right class of operator for the tasks
		required (logical operators for bzip2 and floating-point operators for LivermoreC)
		but the code still requires transformation to fit the individual available
		operations.
	}
	\label{CaseStudiesCompilationRateGraphSplit}
\end{figure*}

We apply \name{} to four accelerators presented in section~\ref{ExistingDSASection},
comparing to three other rewriting strategies.
\Cref{CaseStudiesCompilationRateGraph} shows that \name{} increases the number of loops that these
CGRAs can support by a factor of 2.2$\times$. \Cref{CaseStudiesCompilationRateGraphSplit} gives details split by benchmark suite for each accelerator.

\subsubsection{DSAGen}
Figure~\ref{CaseStudiesCompilationRateGraph} shows
that using \name{} increases the number of loops that
can be supported on the CCA and Maeri architectures by a factor
of 2.2$\times$ and 1.6$\times$ respectively. Maeri does particularly well on LivermoreC (figure~\ref{CaseStudiesCompilationRateGraphSplit}),
especially once equality saturation is used, because of the workload's heavy use of
floating-point operations, though it is less suited to Bzip2 than CCA because of Maeri's lack of boolean arithmetic.

LLVM performs well on the CCA architecture as it has a more comprehensive set of rewrite
rules than have been implemented in \name{}, and on the CCA architecture, the
canonicalization rules it uses are appropriately targeted.  Nevertheless,
\name{} outperforms it due to more comprehensive exploration of the rewrite
space.

This case study, on non-CGRA architectures,
reveals the generality of \name: while
we do not claim that this comprehensively
demonstrates that our rewriter
can compile to different architectures (as we still rely
on the OpenCGRA backend), it does
demonstrate that \name{} may be applicable to more
diverse computation models than CGRAs.

\subsubsection{REVAMP}
We implement REVAMP in the OpenCGRA framework and compile each
of our benchmarks to it (\cref{CaseStudiesCompilationRateGraph}).
\name{}
increases the number of loops that can be supported on
this CGRA by 15\%, consistently across different workloads (\cref{CaseStudiesCompilationRateGraphSplit}).

This increase is small
because REVAMP's example already supports almost all
the required operations for non-floating point code.  We will
see in other examples that \name{} becomes more important as the domain
becomes more restricted.

\subsubsection{SC-CGRA}
Figure~\ref{CaseStudiesCompilationRateGraph} shows
\name{} increases the number of loops
that can be supported by a factor of 5.2$\times$.

This case study demonstrates \name{}
is not only relevant within heterogeneous
fabrics: if a homogeneous CGRA lacks
operations that compilers typically
assume to be available, \name's methods may still be necessary to generate
working code. Bzip2 in particular (\cref{CaseStudiesCompilationRateGraphSplit}) more than doubles the amount of targeted code once FlexC's equality saturation is used, compared to greedy-only, because otherwise it gets stuck in local minima and fails to explore the space enough to find a match.

\subsection{Compilation Rate: Architectures Specialized for Loops}
\label{SpecializedLoopsSection}
We demonstrate that the rewriting technique used by \name{}
is applicable to many different specialized
CGRAs within a varied design space.

\begin{figure}
	\includegraphics[width=\columnwidth]{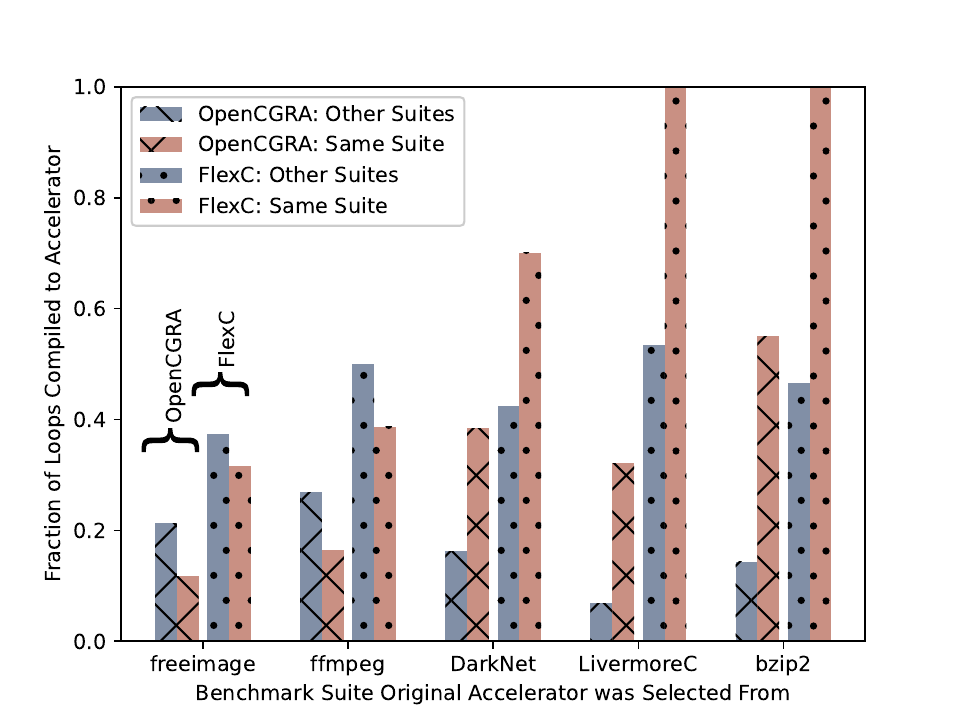}
	\caption{
		Using accelerators designed for individual
		loops in each benchmark suite, how much code
		in the same suite (red) and other suites (blue)
		can be compiled to these accelerators.
		FlexC increase the compilation rate
		by a factor of 2.3$\times$ in the same suite and 2.9$\times$
		between suites.
	}
	\label{CompileRateFigure}
\end{figure}

Using ~300 randomly selected loops in our benchmark suite,
we first build a heterogeneous
CGRA designed for that loop in particular.  We run \name{} across
the other loops in the benchmark suite and measure which loops can
and cannot be compiled.  Figure~\ref{CompileRateFigure} shows what fraction
of loops can be compiled, making a distinction between loops that
are in the same suite (so are often more likely to share the
same class of operation) and loops from different domains.

\name{} improves the applicability of the accelerators, both
within the domain they were designed for by a factor
of 2.3$\times$, and between domains by a factor of 2.9$\times$,
demonstrating the applicability of \name{} to many different
types of heterogeneous CGRA\@. In some cases, a typical accelerator for a loop in one benchmark will actually do better on the other workloads (e,g, for freeimage and ffmpeg). This is because freeimage and ffmpeg are highly diverse, and so an accelerator designed for one loop is less likely to match others in the same diverse benchmark.

\begin{figure}
	\includegraphics[width=\columnwidth]{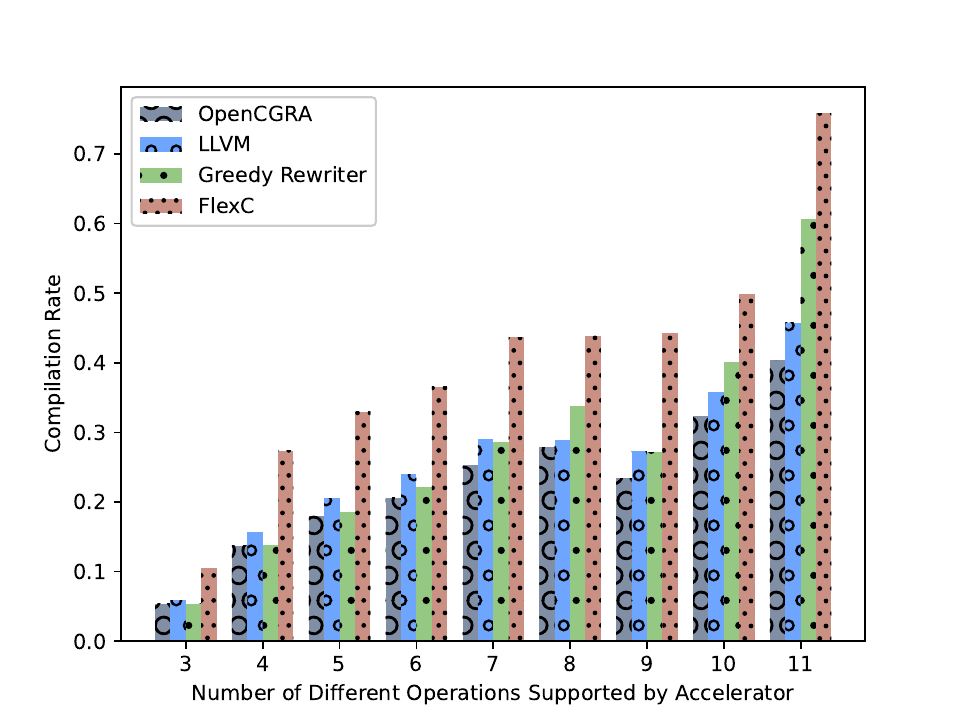}
	\caption{
		How the number of different operations in a CGRA influences compilation
		rate. \name{} performs consistently across many
		levels of generality, from very specialized accelerators with
		few operators to much more generic accelerators with many
		different operators. Equality saturation is
		most important for more specialized architectures.
	}
	\label{CompileRateBySize}
\end{figure}

Figure~\ref{CompileRateBySize} shows \name{} supports
CGRAs across a wide range of specializations, from very
specialized CGRAs with only a few operators to
complex heterogeneous CGRAs.
For architectures with fewer operations,
equality saturation is more important, as there are
fewer paths to a valid rewrite.

\subsubsection{Speedups}
\label{SpeedupSection}
This section demonstrates that rewriting code
in ways that at first-glance are inefficient
can result in speedup by enabling accelerator utilization.
Compiling to CGRA implementations typically improves
performance \textit{and} reduces power usage.
We consider speedup in this evaluation.
In line with other CGRA work, we consider speedup
in the case that loops are executing large numbers
of iterations, so one-time overheads like offloading costs
for loosely coupled accelerators are ignored.

\begin{figure}
	\includegraphics[width=\columnwidth]{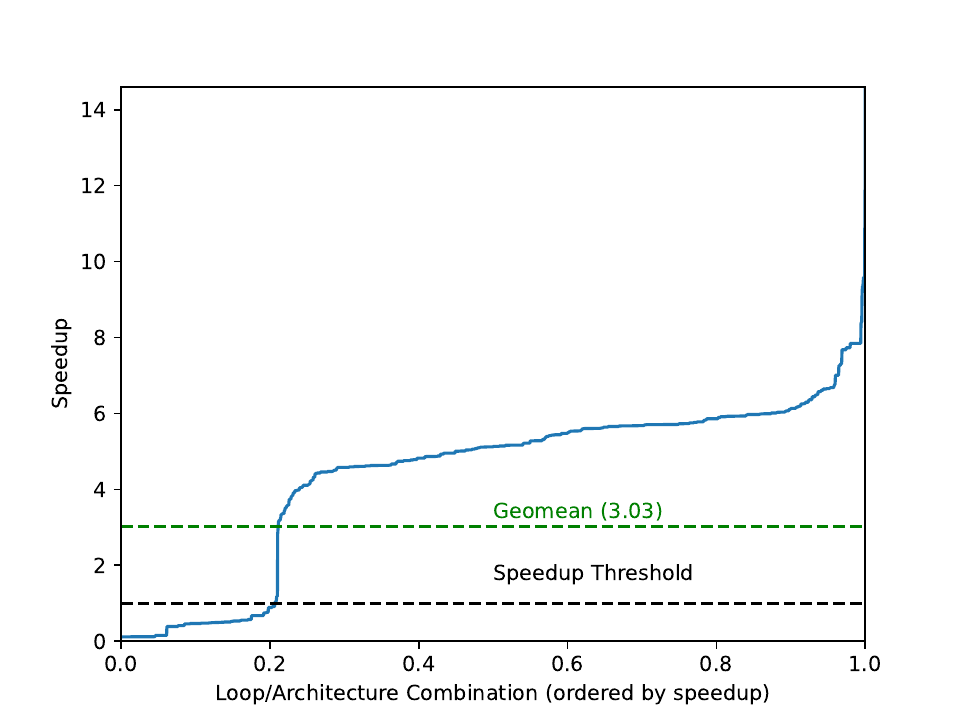}
	\caption{
		Speedup achieved by rewriting applications to run on a
		low-power CGRA vs running on a comparable low-power CPU\@.
		We achieve 3$\times$ geomean over running on a CPU\@.}
	\label{SpeedupsCDF}
\end{figure}

We compare two systems with similar specifications.
For a CGRA system,
we take architectural parameters for ADRES~\cite{Mei2003}, a 6x6 CGRA which
we clock at 200~MHz.
We use
the initialization interval
to obtain performance estimates
for the CGRA\@.
To obtain a realistic CPU baseline, we execute  the loops on
an Arm A5 running at 500~MHz using an Analog Devices SC-589EZKit development
board~\cite{AnalogDevices2018} and methodology for generating
inputs from Exebench~\cite{Armengol-Estape2022}.
Speedups are shown in figure~\ref{SpeedupsCDF}, showing
a geomean performance improvement of 3$\times$, demonstrating
that \name{}'s rewrite rules are not only effective
in enabling targeting of CGRAs, but also in achieving
speedup on them.


\subsection{Existing Domain-Specific Accelerators: End-to-End Evaluation}
We demonstrate that \name{} also performs well on
well-known and computationally important kernels.
To do this, we take the OpenCGRA benchmark suite~\cite{tan2020}, along
with the LivermoreC benchmark suite previously explored.
We use the same setup as in section~\ref{SpeedupSection}.

\begin{figure}
	\includegraphics[width=0.45\textwidth]{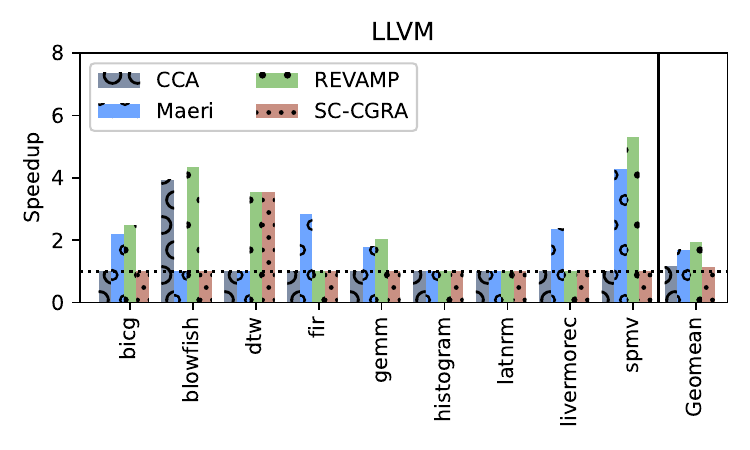}
	\includegraphics[width=0.45\textwidth]{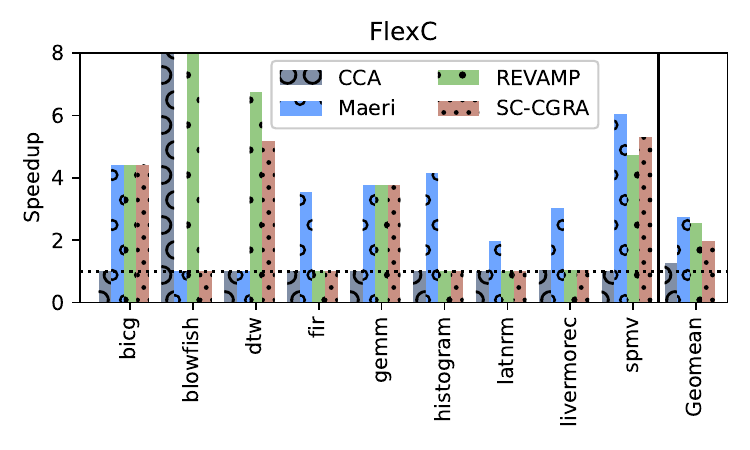}
	\caption[In-Context Speedups]{
		Speedups using the OpenCGRA benchmark suite and the Livermore C
		benchmark suite, comparing various CGRA architectures to
		an Arm Cortex-A5.  Benchmarks that were unsupported by any
		architecture/compiler pairs have been omitted\footnotemark{}.  
		The top figure shows the speedup achieved using the LLVM rewriter to target each CGRA, and the bottom figure shows speedup via \name{}.
	}
	\label{InContextFigure}
\end{figure}

The results are shown in Figure~\ref{InContextFigure}.
Compared to running on an Arm Cortex-A5, \name{} achieves
a speedup on 2.0$\times$ across all applications.  This compares
to the LLVM rewriter, which is only able to extract 1.5$\times$ performance increase
across all applications.

\footnotetext{We omitted ADPCM Encoder/Decoder as OpenCGRA is unable to compile it to any accelerators due to size, and
			Conv, FFT, MVT and Relu due to presence of divide operations that cannot be eliminated, as none of our case-study accelerators support
		divide operations. }

\subsection{Using Different Rulesets}
\name{} provides a generic rewriting framework that
can be applied to many different rulesets.  These rulesets
may be flagged by the programmer as valid for particular loops,
or valid for a particular program.

\begin{figure}
	\includegraphics[width=\columnwidth]{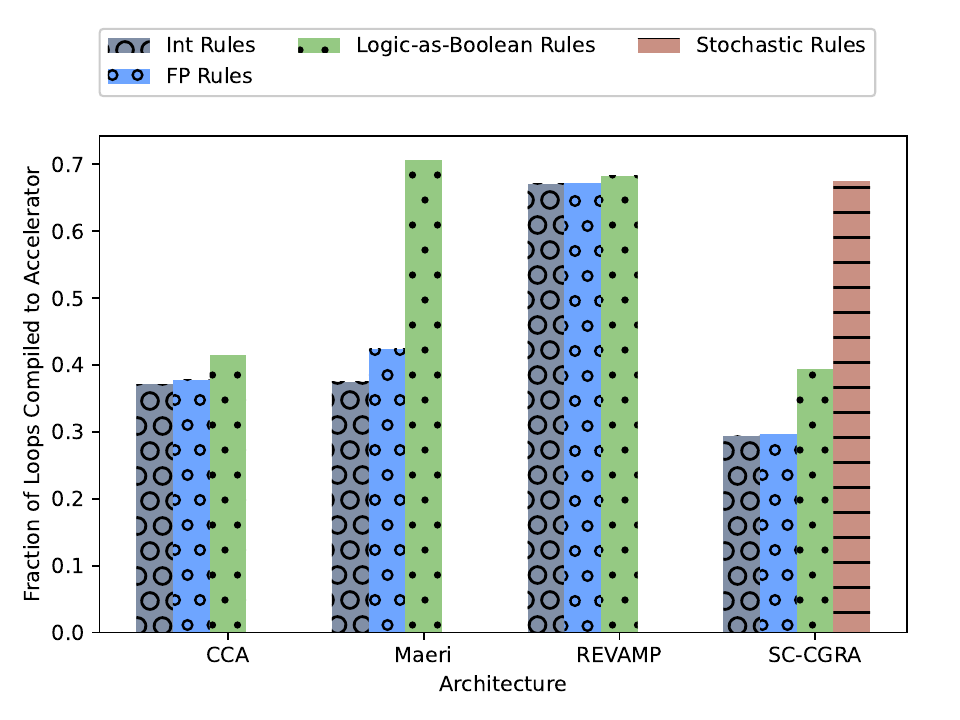}
	\caption{
		Comparing how different sets of rewrite rules
		improve the code coverage of an accelerator.
		All rulesets are run with the int ruleset.
		The stochastic computing rules are only applied
		to SC-CGRA as they require specialized hardware
		support not available in other accelerators.
	}
	\label{CompilationRateUnderDifferentRulesets}
\end{figure}

We inspect four different rulesets here (covered in more detail
in Section~\ref{RulesetsSection}),
an
integer ruleset, derived of rules that may always be applied,
a floating-point ruleset, derived of rules that may be applied
under assumptions such as \texttt{-ffast-math},
a logical operations-as-binary operations ruleset that
can be used to provide greater flexibility of rewrites involving
logical operators
and a stochastic computing ruleset that enables typical stochastic
computing transformations.
These secondary rulesets can be activated by the programmer
using a flag.
Figure~\ref{CompilationRateUnderDifferentRulesets} shows
how these different rulesets provide different compilation performance.
Rulesets are run in combination with the int ruleset as it contains
many \textit{enabling} rewrites for the specialized rewrites
in the other rulesets.
We can see that determining which rulesets are useful is
architecture-specific.  For example, Maeri benefits
a lot from the logic-as-boolean ruleset, as it does
not have logic operators, while CCA benefits
from the stochastic rules as it does not have multipliers.

\subsection{Most Frequently Applied Rewrite Rules}
Part of the power of \name{} is that the rewrite rules
that need to be applied vary by architecture.  By using
equality saturation, \name{} is able to use one standard
set of rules for all architectures and apply the relevant
rules in each case.  Table~\ref{WhichRulesTable} shows
the most frequently applied rules for the CCA and Maeri
architectures (when compiled using the integer and floating
point rulesets): two architectures with nearly disjoint sets
of operators.

\begin{table}
		\centering
	\begin{tabular}{|c|c|}
		\hline
		& CCA \\
		\hline
		1 & \verb|x * 2 => x << 1| \\
		2 & \verb|x * 4 => x << 2| \\
		3 & \verb|x * 1 => x| \\
		4 & \verb|-x (Floating Point) => x + 2^32 (Int)|  \\
		\hline
		\hline
		  & Maeri \\
		  \hline
 1 & \verb|x * 1 => x| \\
 2 & \verb|x << 1 => x * 2| \\
 3 & \verb|x << y => mul(x, load(csel(y > 32, 33, y)))| \\
4 &  \verb|x - y => x + -y| \\
		\hline
		\hline
  & REVAMP \\
  \hline
1 & \verb|x * 1 => x| \\
2 & \verb|-x (Floating Point) => x + 2^32 (Int)|  \\
3 & \verb|x  / 2 => x >> 1| \\
4 & \verb|x / 8 => x >> 3| \\
\hline\hline
  & SC-CGRA \\
  \hline
1 & \verb|x * y => x & y| \\
2 & \verb|x * y => ISC(x, y)| \\
3 & \verb|x * 1 => x| \\
4 & \verb|-x (Floating Point) => x + 2^32 (Int)|  \\
\hline
\hline
	\end{tabular}
	\caption{The most commonly applied rules for
		each architecture.  We omit
		LLVM-specific rewrites for SC-CGRA\@.
		As the CCA and Maeri
		provide nearly disjoint operators, they
		are examples of
		the need for rewrites to apply bidirectionally.
	}
	\label{WhichRulesTable}
\end{table}

\begin{figure}
	\includegraphics[width=\columnwidth]{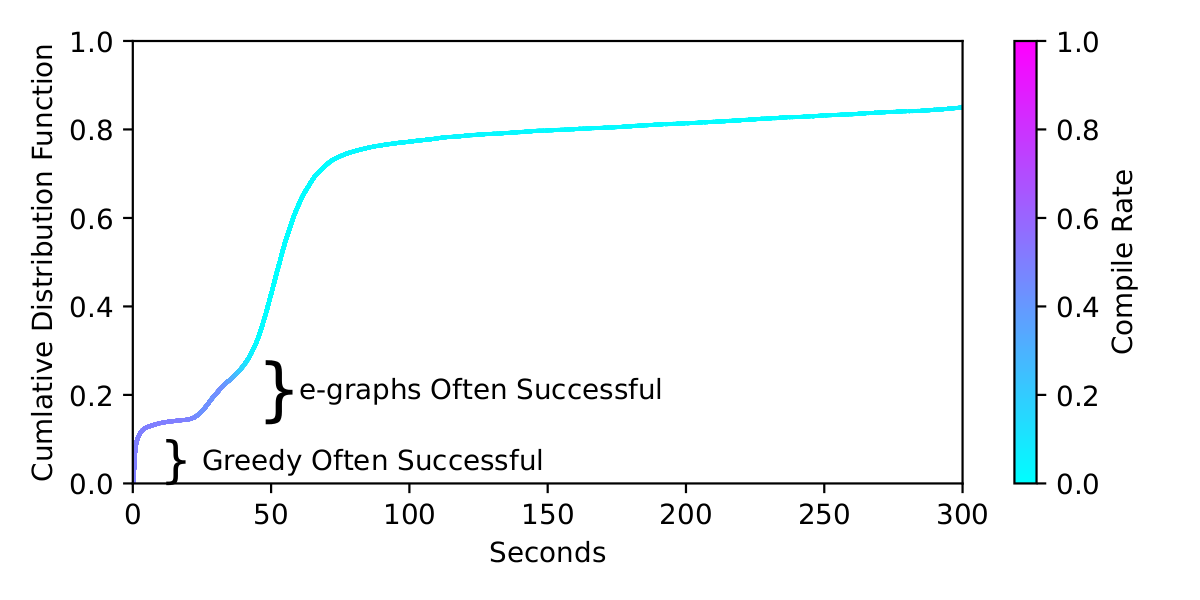}
	\caption{
		Time to schedule code on a CGRA using
		\name{} and OpenCGRA\@.  We cut-off rewriting at
		300~s to avoid excessive exploration.  After 60~s, the compilation rate is very low,
		so \name{} is not missing many compilations
		at longer timeouts.
	}
	\label{CompilationTimeFigure}
\end{figure}

\subsection{Compile Time}
A challenge with Equality Saturation is in keeping the search-space
manageably sized, as e-graphs can grow rapidly, causing excessive compile
times and resource usage~\cite{Koehler2022, kourta2022caviar}.
We avoid these issues in \name{} by limiting the number of explorative iterations, still finding good solutions in many cases.

Figure~\ref{CompilationTimeFigure} shows the time taken
by \name{} to rewrite and schedule the DFG\@.  We use
a cutoff time of 300~s to avoid exploring the rewrite space
fruitlessly --- we can see that the rate of successful compilations drops off rapidly after 60~s, followed first by a large number of early terminations without successful scheduling (most likely due to reaching saturation, iteration or node limit), then by stagnation in progress for infeasibly large search spaces.  These compile
times are fast enough that more exhaustive CGRA schedulers
will be able to incorporate this strategy within the existing
order of magnitude of compilation time. For example
Beidas and Anderson report ILP compile times with a geomean
of 60~s~\cite{Beidas2022}.

We can also see the effect of using a greedy rewriter
as a preliminary step here.  In 10\% of cases,
\name{} is able to rely on the greedy rewriter
and find a compiling loop rapidly.  We can further see that
when \name{} uses equality saturation, it is more
successful early on in the exploration.







\section{Related Work}
\subsection{Existing CGRAs}
Research on CGGRs has been extensive~\cite{Liu2019a,Martin2022}.
Older CGRAs~\cite{Mei2003,Hauser1997,Govindaraju2012}
tend to provide a homogeneous grid of PEs with
a programmable interconnect.
However, the design-space of CGRAs is immense,
including heterogeneous on-chip networks~\cite{Zhou2013,Wijerathne2022,Madhu2015,Karunaratne2017,Adhi2022},
decouplings of memory and compute~\cite{Wijerathne2019,Weng2020},
unifying memories~\cite{Dai2022},
and various techniques to specialize
PEs~\cite{Zhao2017a,Ansaloni2011,Park2009,Melchert2021,Willsey2019,Gobieski2021,Stojilovic2013,Byun2022}.
Toolchains to enable development~\cite{Chin2017,Tan2021a,Tan2021,Liu2015b,Podobas2020,Silva2020,Guo2020,Wijerathne2022a}
and aid design~\cite{Bandara2022,Dobrich2011,Weng2021}
mean that is is relatively easy to design and build
a domain-specific CGRA\@.

\subsubsection{Domain-Specific CGRAs}
Domain-specific CGRAs exist for
neural networks~\cite{Geng2020,SambaNovaSystems2021,Anwar2014,Lee2021a},
scientific kernels~\cite{Charitopoulos2021,Denkinger2022,Byun2022,Ebrahimi2021,Prasad2021,Liu2019d},
approximate computing~\cite{Akbari2018,Wang2022},
stencil computations~\cite{Tithi2021},
HPC~\cite{Madhu2015}, multimedia~\cite{Yang2012,Mei2008,Nguyen2022} and
streaming applications~\cite{Li2022a}.

\subsubsection{Industrial CGRAs}
Xilinx's ACAP provides a CGRA-like model of computation~\cite{Xilinx2022} and uses
an MLIR-based toolchain~\cite{Xilinx2022a}.
Samsung have designed the SLP-URP~\cite{Kim2012} for low-power medical
use-cases.  Smaller companies such as Wave Computing~\cite{Nicol2017} and SambaNova Systems~\cite{SambaNovaSystems2021} and Recore Systems~\cite{Heysters2005} also
involved in CGRA-design.
Large-scale research projects
of producing real hardware~\cite{Pal2020,Gomony2023} also
use CGRAs.

\subsection{Compiling for CGRAs}
Numerous authors  address compiling branches~\cite{Yuan2021,Karunaratne2019,Balasubramanian2018,Wood2017,Nguyen2021,Gobieski2022},
nested loops~\cite{Karunaratne2018,Wood2017,Ruschke2016}, scheduling
of large loops~\cite{Kojima2022,Hamzeh2012,Egger2017,Ohwada2022,Kou2022}
and irregular memory accesses~\cite{Gobieski2022}.
Compilation time is relevant in many fields~\cite{Lee2021,Wirsch2021}
and has been addressed both with faster algorithms~\cite{Lee2021} and
hardware acceleration~\cite{Vieira2021,Canesche2021}.

CGRA scheduling can be done with
binary decision diagrams~\cite{Beidas2022},
the polyhedral model~\cite{Madhu2015,Liu2013},
SAT solvers~\cite{Miyasaka2021,Nowatzki2013} and
ILP models~\cite{Walker2019,Chaudhuri2017,Yoon2008,Mu2021}.
Heuristic approaches can use
information from failed placements~\cite{Balasubramanian2022,Zhao2020},
rewrite rules to simplify routing~\cite{Kim2014}, sharing
information between placement and routing phases~\cite{Dave2018}
and integration of hardware features within the compiler model~\cite{Kou2020,Yang2017}.
Machine learning can automate many
approaches~\cite{Liu2019c,Kao2020,Chang2022,Kou2022,Zhuang2022}.

DSLs can be used to simplify
the compilation processes, enabling
greater parallelism~\cite{Prabhakar2017,Koeplinger2018,Keryell2021,Liu2021b,Bahr2020,Zhang2021} but do not
totally eliminate compiler challenges~\cite{Feldman2021}.  API interfaces can eliminate compiler challenges, but also eliminate
the flexibility of CGRAs~\cite{Lopes2016,Rakossy2015}.

\subsection{Compiling for Hardware Accelerators}
Compiling for API programmable accelerators has been
explored using equality saturation~\cite{huang2022-flexible-mapping}
and program synthesis~\cite{Woodruff2022}.
Equality saturation has also been used to optimize tensor
programs for tensor accelerators~\cite{smith2021pure}.
Similarly, externalizing rewrite rules to
enable programs to run efficiently on hardware accelerators~\cite{Ikarashi2022}.
While these approaches could target CGRAs behind API
interfaces, they do not support compiling to CGRAs directly --- and
so lose flexiblity.
Idiomatic compilation~\cite{Ginsbach2020} has been used
to target closely-related spatial accelerators~\cite{Weng2022}.

\subsection{Equality Saturation}
Equality saturation \cite{tate2009-equality-saturation, Willsey2021} has been used for a range of tasks, including:
optimization and translation validation of Java bytecode and LLVM programs \cite{stepp2011equality}, 
improving accuracy of floating point expressions \cite{panchekha2015automatically}, 
synthesizing structured CAD models \cite{nandi2020synthesizing}, 
optimizing linear algebra expressions \cite{wang2020spores}, 
tensor graph superoptimization \cite{yang2021equality}, 
vectorization for digital signal processors \cite{vanhattum2021vectorization}, 
optimizing integer multiplication on FPGAs \cite{ustun2022impress}, 
hardware datapath optimization \cite{coward2022automatic}.

A proposed DSLs to Accelerators (D2A) methodology \cite{smith2021pure, huang2022-flexible-mapping} uses equality saturation for optimization and hardware mapping
of DSLs\@.
This paper aligns with the \emph{flexible matching} idea from the D2A methodology, but considers mapping arbitrary C code to CGRAs to address the CGRA domain-restriction problem,
and evaluates the difference between equality saturation and greedy rewriting.

While equality saturation is a powerful rewriting technique that addresses limitations of greedy rewriting, scaling to long rewrite sequences is limited as the e-graph grows quickly.
The Pulsing Caviar mechanism~\cite{kourta2022caviar} was evaluated on arithmetic expressions to balance exploration and exploitation, and compared to greedy rewriting for this purpose.
Sketch-guiding~\cite{Koehler2022} is another recently proposed semi-automated technique to improve scaling of equality saturation.

\section{Conclusion}
We introduce \name{}, a compiler for domain-specific CGRAs that
addresses the domain-restriction problem: where CGRAs
that have been designed for a particular domain are hard to apply
to software outside that domain.
\name{} uses equality-saturation to rewrite software
from different domains so it can run on hardware not designed
for it.
\name{} increases the number loops that can be supported
by a factor of 2.2$\times$ over existing CGRA compilers and
enables acceleration of loops leading to a geomean speedup of 2.1$\times$.

\name{} demonstrates the potential that rewriting software
to match novel hardware has: the techniques developed
here are applicable to other kinds of accelerators with
programmable networks.
We present the first study that characterizes how
different decisions surrounding heterogeneity effect
the fraction of code supported by an accelerator, showing
that the more specialized an accelerator is, the more
important \name{} is.
\name{} opens up new development possibilities
by promising that even if software requirements change in
a heartbeat, accelerators with a large sunk-cost
can still be applied.

\balance

\bibliographystyle{ieeetr}
\bibliography{refs}

\end{document}